\documentstyle[psfig]{mn2e}

\def\gs{\mathrel{\raise0.35ex\hbox{$\scriptstyle >$}\kern-0.6em
\lower0.40ex\hbox{{$\scriptstyle \sim$}}}}
\def\ls{\mathrel{\raise0.35ex\hbox{$\scriptstyle <$}\kern-0.6em
\lower0.40ex\hbox{{$\scriptstyle \sim$}}}}
\def\ls{\mathrel{\hbox{\rlap{\hbox{\lower4pt\hbox{$\sim$}}}\hbox{$<$}}}}
\def\gs{\mathrel{\hbox{\rlap{\hbox{\lower4pt\hbox{$\sim$}}}\hbox{$>$}}}}

\def\msun{{\rm\,M_\odot}}

\title[The HIPASS Catalogue: III - Optical Counterparts \& Isolated Dark Galaxies]
      {The HIPASS Catalogue: III - Optical Counterparts \& Isolated Dark Galaxies}
\author[M.~T.~Doyle et al.]
       {M.~T.~Doyle,$^{1,\star}$ M.~J.~Drinkwater,$^{1}$ D.~J.~Rohde,$^{1}$ 
K.~A.~Pimbblet,$^{1}$ M.~Read,$^{2}$ \and
M.~J.~Meyer,$^{3,4}$ M.~A.~Zwaan,$^{5}$ E.~Ryan-Weber,$^{3,6}$ J.~Stevens,$^{3}$ B.~S.~Koribalski,$^{7}$
 \and
R.~L.~Webster,$^{3}$ L.~Staveley-Smith,$^{7}$ D.~G.~Barnes,$^{3}$ M.~Howlett,$^{8}$ 
V.~A.~Kilborn,$^{8\,9}$
\and
M.~Waugh,$^{3}$ M.~J.~Pierce,$^{8}$ R.~Bhathal,$^{10}$ W.~J.~G.~de Blok,$^{11}$ 
M.~J.~Disney,$^{11}$ \and
R.~D.~Ekers,$^{7}$ K.~C.~Freeman,$^{12}$ D.~A.~Garcia,$^{11}$ B.~K.~Gibson,$^{8}$ 
J.~Harnett,$^{13}$ \and
P.~A.~Henning,$^{14}$ H.~Jerjen,$^{12}$ M.~J.~Kesteven,$^{7}$ P.~M.~Knezek,$^{15}$ 
 \and
S.~Mader,$^{7}$ M.~Marquarding,$^{7}$ R.~F.~Minchin,$^{11}$ J.~O'Brien,$^{12}$ 
T.~Oosterloo,$^{16}$ \and
R.~M.~Price,$^{14}$ M.~E.~Putman,$^{17}$ S.~D.~Ryder,$^{18}$ E.~M.~Sadler,$^{19}$ 
I.~M.~Stewart,$^{20}$ \and
F.~Stootman$^{10}$ and A.~E.~Wright $^{7}$
        \vspace*{1mm}\\
	$^{\star}$ mtdoyle@physics.uq.edu.au \\
  $^{1}$ Department of Physics, University of Queensland, Brisbane, QLD 4072 Australia\\
	$^{2}$ WFAU, Institute for Astronomy, Royal Observatory,	Blackford Hill, Edinburgh, EH9 3HJ, UK\\
	$^{3}$ School of Physics, University of Melbourne,	VIC 3010, Australia\\
	$^{4}$ Space Telescope Science Institute, 3700 San Martin Drive, Baltimore MD 21218, USA\\
	$^{5}$ European Southern Observatory, Karl-Schwarzschild-Str.2, 85748 Garching b. Munchen, Germany\\
	$^{6}$ Institute of Astronomy, University of Cambridge, Madingly Road,	Cambridge, CB3 0HA, UK\\
	$^{7}$ Australia Telescope National Facility, CSIRO, P.O. Box 76, Epping, NSW 1710, Australia\\
	$^{8}$ Centre for Astrophysics and Supercomputing, 	Swinburne University of Technology, P.O. Box 218, Hawthorn, VIC 3122 Australia\\
	$^{9}$ Jodrell Bank Observatory, University of Manchester, Macclesfield,Cheshire, SK11 9DL, UK\\
	$^{10}$ Department of Physics, University of Western Sydney Macarthur, P.O.Box 555, Campbelltown, NSW 2560, Australia\\
	$^{11}$ School of Physics \& Astronomy, Cardiff University, The Parade, Cardiff CF24 3YB, UK\\
	$^{12}$ Research School of Astronomy \& Astrophysics, Mount Stromlo Observatory,Cotter Road, Weston, ACT 2611, Australia\\
	$^{13}$ University of Technology Sydney, Broadway	NSW 2007, Australia\\
	$^{14}$ Institute for Astrophysics, University	of New Mexico, 800 Yale Blvd, NE, Albuquerque, NM 87131, USA\\
	$^{15}$ WIYN, Inc. 950 North Cherry Avenue, Tucson, AZ, 85719 USA\\
	$^{16}$ ASTRON, P.O. Box 2, 7990 AA Dwingeloo, The Netherlands\\
	$^{17}$ CASA, University of Colorado, Boulder, CO 80309-0389, USA\\
	$^{18}$ Anglo-Australian Observatory, P.O.Box 296, Epping, NSW 1710, Australia\\
	$^{19}$ School of Physics, University of Sydney, NSW 2006, Australia\\
	$^{20}$ Department of Physics \& Astronomy, University of Leicester, Leicester LE1 7RH, UK}

\date{\fbox{\sc Draft: \today\ --- Do Not Distribute}}
\pagerange{000--000}

\begin{document}

\maketitle

\begin{abstract}

We present the largest catalogue to date of optical counterparts for HI radio-selected galaxies, \textsc{Hopcat}. Of the 4315 HI radio-detected sources from the HI Parkes All Sky Survey \textsc{(Hipass)} catalogue, we find optical counterparts for 3618 (84\%) galaxies. Of these, 1798 (42\%) have confirmed optical velocities and 848 (20\%) are single matches without confirmed velocities. Some galaxy matches are members of galaxy groups. From these multiple galaxy matches, 714 (16\%) have confirmed optical velocities and a further 258 (6\%) galaxies are without confirmed velocities. For 481 (11\%), multiple galaxies are present but no single optical counterpart can be chosen and 216 (5\%) have no obvious optical galaxy present. Most of these `blank fields' are in crowded fields along the Galactic plane or have high extinctions.

Isolated `Dark galaxy' candidates are investigated using an extinction cut of $A_{B_j}$ $<$ 1 mag and the blank fields category. Of the 3692 galaxies with an $A_{B_j}$ extinction $<$ 1 mag, only 13 are also blank fields. Of these, 12 are eliminated either with follow-up Parkes observations or are in crowded fields. The remaining one has a low surface brightness  optical counterpart. Hence, no isolated optically dark galaxies have been found within the limits of the \textsc{Hipass} survey.
\end{abstract}


\section{Introduction}
  \label{Sec:Introduction}
The blind HI Parkes All Sky Survey (\textsc{Hipass}), on the Parkes Radio Telescope was completed in 2000.  This survey covers the whole of the southern sky  up to Dec = +2$^{o}$ \footnote{A northern extension to +25$^{o}$ has also been observed and the resulting catalogue is in preparation.}.  The \textsc{Hipass} catalogue, \textsc{Hicat}, represents the largest HI-selected 
catalogue at this time and is presented  in a series of papers, this one being the third. In Paper I, Meyer et al.\ (2004) describe the selection procedure, global sample properties and the catalogue. In Paper II, Zwaan et al.\ (2004), describe the completeness and reliability of \textsc{Hicat}. Though the survey and catalogue are developed to be free of optical selection bias, it is nevertheless important for several scientific applications to have accompanying optical data.
This paper details the process taken to find the optical counterparts for \textsc{Hicat} sources. 

Two of the (many) motivations for the \textsc{Hipass} survey are to investigate low surface brightness (LSB) and dark galaxies. For LSB galaxies, the goal is to provide a large sample sensitive to LSB galaxies previously undetected by optical methods, such as Malin 1 type objects (Bothun et al.\ 1987). By using an HI radio selected sample, any study of LSB galaxies will have minimal optical bias. It was originally thought that a large population of LSB galaxies would be revealed by HI surveys like \textsc{Hipass} (Disney 1976), but they have not been detected in the recent surveys (e.g. Minchin et al.\ 2004).

At the limit of low surface brightness are ``dark galaxies'' with no detected optical emission.  It is important to separate true dark galaxies from gas clouds which are directly associated with optical galaxies, often through tidal interactions. For the purposes of this paper we define a dark galaxy as any HI source that contains gas (and dark matter) but no detectable stars, and is sufficiently far away from other galaxies, groups or clusters such that a tidal origin can be excluded.  One of the most striking examples of associated HI gas clouds is the Leo ring, a massive cloud of tidally-disrupted gas in the Leo galaxy group (Schneider et al.\ 1983). More recently, early studies of the HIPASS survey have detected associated gas clouds in
the NGC~2442 group (Ryder et al.\ 2001) and in the Magellanic Cloud-Milky Way system (Kilborn et al.\ 2000).

There is a range of theoretical opinion about the existence of dark galaxies. Verde et al.\ (2002) argue that a large fraction of low-mass dark matter halos would form stable gas disks on contraction and thus not exhibit star formation, whereas Taylor \& Webster (2005) conclude that HI clouds cannot exist in equilibrium with the local universe without becoming unstable to star formation.

There are few, if any, true dark galaxy detections. Some sources, initially described as dark galaxy detections, have since been identified with optical galaxies such as the protogalaxy HI\,1225+01 (re-detected here as HIPASS J1227+01) in the Virgo cluster (Giovanelli \& Haynes 1989), later detected in the optical by Salzer et al.\ (1991). Several authors have placed upper limits on the numbers of dark galaxies in previous HI surveys.  Fisher \& Tully (1981) found no evidence for dark galaxies in a survey of 153 square degrees and estimated that the total mass of any dark HI clouds ($107\msun<M_{HI}<10^{10}\msun$) was less than 6 per cent that in normal galaxies. More recently Briggs (1990) showed that the space density of dark galaxies is less than 1 percent that of normal galaxies for masses $M_{HI}<10^{8}\msun$, and the the Arecibo surveys by Zwaan et al.\ (1997) also failed to detect any dark galaxies in the range ($10^{7.5}\msun<M_{HI}<10^{10}\msun$).

Previous to HIPASS, several blind HI surveys that identify optical counterparts have been carried out, such as the Arecibo HI Sky Survey, AHISS, (Sorar 1994; 66 sources), the Slice Survey (Spitzak \& Schneider 1998; 75 sources) and the Arecibo Dual-Beam Survey, ADBS, (Rosenberg \& Schneider 2000; 265 sources). However \textsc{Hicat} is much larger, containing 4315 HI radio detected objects and covers the whole southern sky from 300 to 12700 km s$^{-1}$.

Newly catalogued galaxies in the \textsc{Hipass} Bright Galaxy Catalogue (BGC, Koribalski et al.\ 2004), are described in Ryan-Weber et al. (2002). Of the 1000 galaxies, 939 have optical counterparts, 4 are high velocity clouds and 57 were deemed to be obscured by dust or confused with stars having galactic latitudes $|$b$|<10$. To date all previous \textsc{Hipass} based studies have conclude that there are no dark galaxies or invisible HI clouds not gravitationally bound to any stellar system present. In using our \textsc{Hicat} optical catalogue, \textsc{Hopcat}, which incorporates the complete Hipass catalogue, a more comprehensive search for isolated dark galaxies in the southern sky is carried out.

In Section 2 we describe the method we use to identify the optical counterparts for the \textsc{Hicat} sources, discuss the input data, how the images are analysed and the process taken to calibrate the magnitudes in the  \itshape{B$_j$}, {R} \upshape \& \itshape {I}\upshape \ bands. In Section 3 we introduce the \textsc{Hipass} optical catalogue, \textsc{Hopcat}. Section 4 lists and analyses the results of the matching process, investigates the optical properties and candidate dark galaxies. A summary of our work is given in Section 5. 

\section{SELECTION OF OPTICAL COUNTERPARTS}

The position of each HI source is used to find the matching optical image. Since \textsc{Hicat} covers a velocity range from 300 to 12700 km s$^{-1}$, finding an optical counterpart, especially those with small peak flux density and at the limits of the survey, poses a variety of challenges. 

In the following sections we describe the preliminary investigations undertaken into finding the best method to optically match the HI radio sources. We detail the image analysis, steps taken to minimize the various problems that arise in matching extended objects and calibrate the resulting magnitudes. We also review what methods and resources are required to cross-check velocities in the matching process and describe the actual matching process in detail.

\subsection{Preliminary Investigations}
  \label{SubSec:PreliminaryInvestigations}
In our preliminary work to identify the best method to find the optical counterparts, we make use of the both the SuperCOSMOS image and data catalogues (Hambly et al.\ 2001a, b \& c). The SuperCOSMOS data catalogue (hereafter SuperCOS data) contains results from image analysis originally optimized for faint star/galaxy classification. This causes problems in the recognition of bright extended objects such as the \textsc{Hipass} galaxies.

When each image from the SuperCOMOS image catalogue (hereafter SuperCOS images) is analysed, ellipses are produced that represent an area around each object above a particular sky intensity for that particular plate. The integrated flux and hence the isophotal magnitude is calculated using these ellipses. However, in the case of the SuperCOS data's ellipses (hereafter SuperCOS ellipses), multiple ellipses can be produced for a single object, especially for extended ones, that segment the area and result in incorrect magnitudes. A different image analysis approach is needed.

\subsection{Optical Image Data\label{SubSec:OpticalImageData} }

Using the results from the preliminary investigations above, the SuperCOS image is optimal for our purposes.  The scanned SuperCOS images have a 10-mm (0.67 arcsec) resolution (Hambly et al.\ 2001b). Each image is required to cover a 7 arcmin radius centred on the \textsc{Hicat} position to allow for any  possible uncertainties in the original coordinates. Using the original positions of the 4315 HI detections from \textsc{Hicat}, 15$\times$15 arcmin \itshape{B$_j$}, {R} \upshape \& \itshape {I}\upshape \ band SuperCOS images, each centred on the \textsc{Hicat} position  are obtained. Independent analysis of each image needs to overcome any segmentation and photometry problems we encounter using the SuperCOS ellipses. 

The SExtractor image analysis package (Bertin \& Arnouts 1996) is used to analyse each SuperCOS image and an SExtractor ellipse catalogue (hereafter SEx ellipses) is produced containing ellipses for each object within the image using the method described in Section \ref{SubSec:PreliminaryInvestigations}.  For the majority of the SuperCOS images, we use SExtractors' default analysis parameters that are optimised for extended and bright objects and that require an higher contrast to detect an object. However, for some objects, the ellipse also includes foreground stars. Therefore a second set of SExtractor parameters, that are more likely to break up detections into smaller objects and that require a much lower contrast to detect fainter objects, are used. A total of 91 images using the second set of parameters are included for the final SEx ellipses.

The magnitudes included in our catalogue are based on SExtractor's \textsc{mag\_auto} defined as `Kron-like elliptical aperture magnitudes' (Bertin \& Arnouts 1996). The magnitude measurements are calculated from the SEx ellipses produced by analysing the $B_j$ band images only and are uncalibrated. Our calibrations of the SExtractor  \itshape{B$_j$}, {R} \upshape \& \itshape {I}\upshape \ magnitudes, using the SuperCOS data's zero point calibration, are described in Section \ref{SubSec:MatchedGalMagCaibration}.

\subsection{Optical spectroscopic Data}

Accurately matching the HI radio detections with their optical counterparts requires a verification method. Using the velocity measurements from \textsc{Hicat} is one way to cross-check the validity of the optical match. Two available resources for velocity cross-checking are NED \footnote{The NASA/IPAC Extragalactic Database (NED) is operated by the Jet Propulsion Laboratory, California Institute of Technology, under contract with the National Aeronautics and Space Administration.} and the 6dF Galaxy  Survey (Wakamatsu et al.\ 2003; 6dFGS; http://www.mso.anu.edu.au/6dFGS/). 

NED is an extragalactic database that draws information from catalogues, surveys and the literature. NED is updated every 2-3 months which poses a problem  for our work as some \textsc{Hipass} velocities are now included. To ensure  that we are not cross-checking \textsc{Hipass} velocities against \textsc{Hipass} and other HI detections, the source of all the NED velocities are checked and only those from optical or high-resolution HI radio observations are included. Since the NED data is taken from various sources, which may vary in quality, a second source of optical velocities is needed.

\begin{figure}
\centerline{\psfig{file=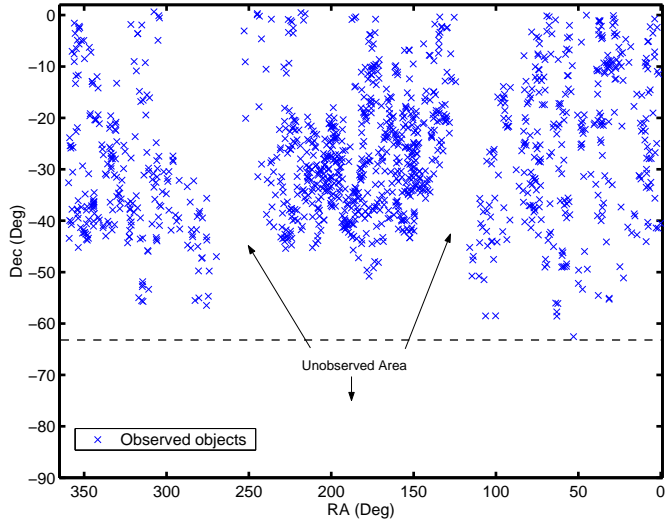,angle=0,width=3.5in}}
  \caption{\small{\textsc{Hicat} sources found in the 6dFGS Pre-release Data for the whole southern sky, showing the 6dFGS's observed and unobserved areas. We use 6dFGS velocities for cross-checking the {\sc Hicat} velocities to verify optical galaxy matches.
}}
  \label{Fig:AllSkyPlot}
\end{figure}

The 6dFGS, an on-going survey of the complete southern sky, is a uniform, high quality dataset that includes redshift measurements. Though 6dFGS is not complete, the sky coverage to date is very useful for our purposes as possible optical counterparts to \textsc{Hipass} sources are included in their observing list. The data we use is not publicly available but are obtained directly from the
6dFGS team and is the latest available at the time. For 6dFGS sky coverage, see Fig. \ref{Fig:AllSkyPlot}.

%
%  Figure 1.
%

\subsection{Magnitude Calibration\label{SubSec:CalibrationMethods} }

The SuperCOS data allows for magnitudes to be calculated as galaxies regardless of whether they are classified as a star or a galaxy (Hambly et al.\ 2001b). This would have been ideal for our purposes, however we are not able to use this data as discussed in Section \ref{SubSec:PreliminaryInvestigations}.

Instead, we use SuperCOS images analysed using the SExtractor image analysis package. This enables us to choose parameters to calculate the area for the SEx ellipses that represents the total galaxy area and hence determine reliable magnitudes. However, this raises problems since the SExtractor magnitudes are not calibrated. To calibrate these magnitudes we use the relationship between SuperCOS data's calibrated and SExtractor's uncalibrated magnitudes. 

\subsubsection{Plate to plate differences }

\begin{figure}
\centerline{\psfig{file=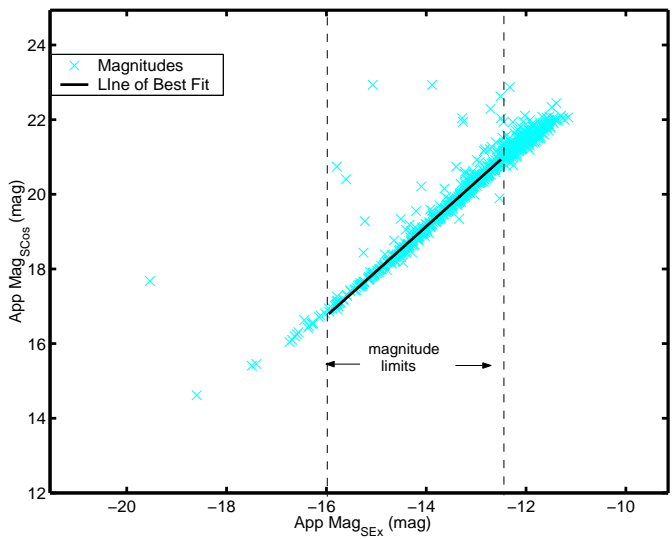,angle=0,width=3.5in}}
  \caption{\small{An example of the relationship between the calibrated SuperCOS data magnitudes and uncalibrated SExtractor magnitudes for every object in a single image. All 4315 images are plotted in this way and by using the slope and intercept from the line of best fit, the zero-point calibration is calculated. We limit the magnitude range used in fitting the line of best fit to exclude object that are too faint or bright, such as very distant objects or foreground stars that may distort the results.
}}
  \label{Fig:SinglePlate2Plate}
\end{figure}

On examination of the images, we find there are variations in the background levels from plate to plate. We use the relationship between the SuperCOS data's calibrated magnitudes and SExtractor's uncalibrated magnitudes to find the zero-point calibration values.  The SuperCOS data's magnitude RMS (zero-point corrected) for \textsc{Hopcat's} magnitude limit are $\sigma_{B_j}$=0.08, $\sigma_{R}$=0.04 and $\sigma_{I}$=0.08 mag (Hambly et al.\ 2001b). For each image the apparent magnitude of every object is plotted and by fitting and using the values from the best line of fit for each plot, a magnitude calibration value for  each plate 
is determined, see Fig. \ref{Fig:SinglePlate2Plate}. The magnitude range of the SuperCOS and SExtracted data has been limited for the line of best fit determination to exclude extremely faint or bright galaxies that may distort the results.

%
%  Figure 2.
%

\subsubsection{Matched Galaxies  \itshape{B$_j$}, {R} \upshape \& \itshape {I}\upshape \ Magnitude Calibration\label{SubSec:MatchedGalMagCaibration}}

\begin{figure}
\centerline{\psfig{file=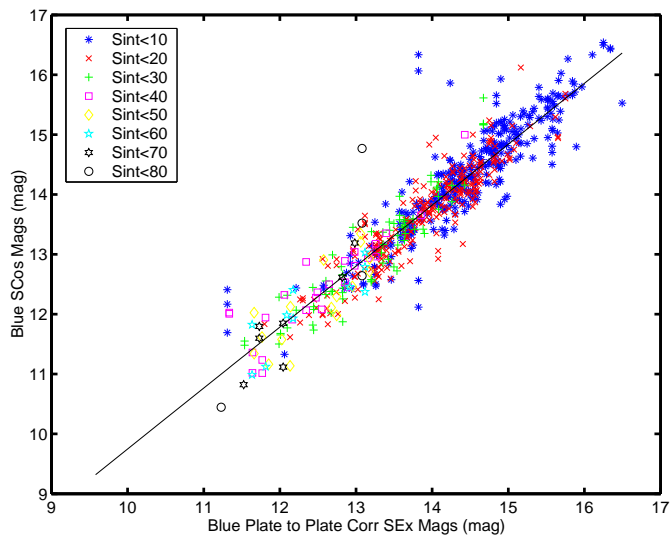,angle=0,width=3.5in}}
  \caption{\small{SuperCOS data's calibrated  \itshape{B$_j$}\upshape \ magnitudes versus SExtractor 
uncalibrated  \itshape{B$_j$}\upshape \ magnitudes. By fitting and using the values from the line 
of best fit, a calibration value for bright objects is determined and the calibrated 
magnitude is calculated for each optically matched galaxy. The corresponding 
\textsc{Hicat} integrated flux strengths (Jy km s$^{-1}$) are listed in the inserted panel.
}}
  \label{Fig:SCosVsSExMagsForCalibration}
\end{figure}

Once the zero-point calibrations for the SuperCOS images are made, we further investigate the calibration for large galaxies by using magnitudes for optically matched galaxies only.  We use the relationship between the calibrated SuperCOS data and the zero-point calibrated SExtractor magnitudes to determine calibrated magnitudes. The SuperCOS data is obtained using the `parent' image deblend parameter to minimise the segmentation problem as previously discussed in Section \ref{SubSec:PreliminaryInvestigations}. Parent image deblends minimise the number of SuperCOS ellipses calculated for a single object. However, on plotting the SuperCOS data and zero-point calibrated SExtractor magnitudes, we find outliers due to the segmentation problem.  We remove these outliers and a line of best fit is determined, Fig. \ref{Fig:SCosVsSExMagsForCalibration}. Using the slope and intercept from the line of best fit and the zero-point calibrated SExtractor magnitudes, Table \ref{Tab:MatchedGalMagCalLOBF}, calibrated \itshape{B$_j$}, {R} \upshape \& \itshape {I}\upshape \ magnitudes are calculated.  

%
%  Table 1.
%
\begin{table}
\begin{center}
\medskip
\caption{\small{Matched galaxy magnitude calibration line of best fit values. 
The slope (m) and intercept (c) are used to calculate the \textsc{Hopcat} calibrated magnitudes.
}}
\begin{tabular}{ccccc}
\hline
Band    & $m$    & $\sigma_m$  &  $c$   &  $\sigma_c$ \\
\hline 
$B_j$       & 1.018  &  0.0002     & -0.438 & 0.0019 \\
$R$       & 0.887  &  0.0019     & 2.2    & 0.01 \\
$I$       & 0.926  &  0.0013     & 2.02    & 0.011 \\
\hline
\end{tabular}
\label{Tab:MatchedGalMagCalLOBF}
\end{center}
\end{table}

The SuperCOS data calculates the  \itshape{B$_j$}, {R} \upshape \& \itshape {I}\upshape \ magnitudes from analysis of \itshape{B$_j$}, {R} \upshape \& \itshape {I}\upshape \ images independently, hence the resulting SuperCOS ellipses for each band vary. However \textsc{Hopcat's} SExtractor magnitudes are calculated using SEx ellipses from  \itshape{B$_j$}\upshape \ band images only. Using the resulting \itshape{B$_j$}\upshape \ based SEx ellipses, the \itshape{R} \upshape \& \itshape {I}\upshape \  magnitudes are calculated using SExtractor analysis of  \itshape{R} \upshape \& \itshape {I}\upshape \  images. As the area of the  \itshape{B$_j$}\upshape \ based SEx ellipses are generally larger than those of the  \itshape{R} \upshape \& \itshape {I}\upshape \ images, the resulting  \itshape{R} \upshape \& \itshape {I}\upshape \ magnitudes will be slightly biased. 

To check our final calibration we compare the \textsc{Hopcat} calibrated magnitudes with published magnitudes from NED and we find a standard deviation of $\sigma=0.6$ mag. This is smaller than the overall spread from our calibration calculations.

%
%  Figure 3.
%

\subsection{The HI Radio-Optical Matching process
\label{SubSec:Radio-OpticalMatching}}

An automated visual interactive program (\textsc{Adric}), where images centred on each \textsc{Hipass} source's position are viewed by several people, has been developed. \textsc{Adric}  utilizes SuperCOS images, SExtractor  \itshape{B$_j$}, {R} \upshape \& \itshape {I}\upshape \ image analysis, and NED and 6dFGS velocities for cross-checking, to reliably match \textsc{Hicat} HI sources with their optical counterparts. 

The matching process consists of downloading  \itshape{B$_j$}, {R} \upshape \& \itshape {I}\upshape \  band SuperCOS images and analysing each one using the 2 sets of image analysis parameters in SExtractor to produce SEx ellipses as discussed in Section \ref{SubSec:OpticalImageData}. These SEx ellipses represent the area of each object and are superimposed onto the image. We use published NED and optical 6dFGS velocities to cross-check the \textsc{Hipass} velocities to confirm the optical match. The NED velocities are checked to ensure that we only use velocities from optical and high-resolution HI radio observations.

When \textsc{Adric} is activated the user views each image, centred on the  \textsc{Hicat} position, with the superimposed SEx ellipses. The original \textsc{Hicat} velocities as well as the 6dFGS and NED velocities are displayed for cross-checking purposes. Each of the 4315 images are matched independently by three people (MTD, MJD \& DJR).  This multiple, independent matching eliminates, as far as possible, biases by any one person in the choosing of galaxy matches. 

We use a 2 digit match category number to quantify the choices made as well as the quality of the resulting magnitudes. The first digit represents the galaxy match choice (Table \ref{Tab:MatchCategoryDescriptions}). The second digit is the quality of the magnitudes based on SEx ellipse coverage. A 0 signifies `good photometry and segmentation'; 1 signifies `poor photometry' where one or more foreground stars are within the SEx ellipse choice; and 2 signifies `poor segmentation' and/or `poor photometry', where the SEx ellipse does not represent the total galaxy area. Both poor photometry and segmentation may effect the magnitude calculation. 

A `velocity match' galaxy (coded as 60, 61 \& 62 in Table \ref{Tab:MatchCategoryDescriptions} \& see Fig. \ref{Fig:Ned6dFMatchedGalaxy}) is chosen when a single galaxy's NED and/or 6dF velocity is within 400km s$^{-1}$ of the \textsc{Hipass} velocity, see section \ref{SubSec:HI Radio-Optical Position and Velocity Separations}. A `good guess' galaxy match (50, 51, 52) is chosen when there is no NED or 6dF velocity but there is a single galaxy within the image that 
could be a match. When there is no agreement on the choice for a galaxy, the image is viewed again
by at least 2 people who then jointly decide on  the correct match category
for the galaxy. 

Note that match categories with first digit = 4 have multiple galaxies present, some with similar velocity values (see Fig. \ref{Fig:MultiMatchedGalaxy}), but the one with the nearest velocity to \textsc{Hicat}'s is chosen. Also if the first digit = 1, one galaxy is chosen without the aid of published velocities but appears to be part of a group.

%
%  Figure 4.
%
\begin{figure}
%\centerline{\psfig{file=Ned6dFMatchScreenShotHipass1060WithText.eps,angle=0,width=3.5in}}
  \caption{\small{(For this figure see MNRAS)Screenshot of the automated visual interactive program (\textsc{Adric}) with NED and 6dF velocity confirmation for the chosen galaxy.}}
  \label{Fig:Ned6dFMatchedGalaxy}
\end{figure}

%
%  Figure 5.
%
\begin{figure}
%\centerline{\psfig{file=NedMultiMatchScreenShotHipass3304WithText.eps,angle=0,width=3.5in}}
  \caption{\small{(For this figure see MNRAS)Screenshot of the automated visual interactive program (\textsc{Adric}) with multiple possible galaxy choices. Here the chosen galaxy is the one with the smallest \textsc{Hipass}-NED velocity difference, 3320km  s$^{-1}$.}
}
\label{Fig:MultiMatchedGalaxy}
\end{figure}

%
%  Table 2.
%
\begin{table*}
\begin{center}
\medskip
\caption{\small{Match choice parameters for the optical counterparts for the \textsc{Hicat} HI detections
}}
\begin{tabular}{ll}
\hline
Match Category Number      & First Digit Description \\
\hline 
Velocity Match & Single velocity match with chosen SEx ellipse \\
(60$^{\dagger}$, 61$^{\dagger}$, 62$^{\dagger}$) & (see Fig. \ref{Fig:Ned6dFMatchedGalaxy})\\
Good Guess & No velocity for SEx ellipse choice \\
(50$^{\dagger}$, 51$^{\dagger}$, 52$^{\dagger}$) & Only significant galaxy in the field \\
 & Galaxy chosen using widths, flux or position (educated guess) \\
Multi Velocity Match & Multiple galaxies within the field with: \\
(40$^{\dagger}$, 41$^{\dagger}$, 42$^{\dagger}$) & $>1$ matching velocity \\
 & $\ge 2$ group members Choose good/best galaxy including merging galaxies \\
 & (see Fig. \ref{Fig:MultiMatchedGalaxy})\\
Blank Field &  No apparent galaxies in field \\
(30) & \\
No Guess & No match \\
(20) & No ellipse chosen \\
     & No similar velocities and there are multiple galaxies in field \\
Multi Good Guess & No velocity for ellipse choice \\
(10$^{\dagger}$, 11$^{\dagger}$, 12$^{\dagger}$) & Multiple galaxies within field with no velocity matches \\
 & $\ge 2$ group members Choose good/best galaxy Include merging galaxies \\
\hline
\end{tabular}

$^{\dagger}$ First digit is the category of the galaxy match.  Second digit is the quality of the resulting magnitude value: 0 = good  photometry and segmentation; 1 = poor photometry ; 2 = poor segmentation.
\label{Tab:MatchCategoryDescriptions}
\end{center}
\end{table*}

\section{The Optical \textsc{Hicat} Catalogue, \textsc{Hopcat}}

Once the optical matches are agreed upon, the data for the Optical \textsc{Hicat} catalogue, \textsc{Hopcat}, are collated. Calibration work is carried out and columns are added from calculations and information such as SuperCOSMOS plate numbers, galactic coordinate, $E(B-V)$ extinctions, galaxy names and morphology. For a  full description of \textsc{Hopcat's } columns see Table \ref{Tab:HopcatColumns}.

The raw data from the interactive visual matching process, \textsc{Adric}, produce uncalibrated magnitudes. As discussed in Section \ref{SubSec:CalibrationMethods} we use the relationship between the calibrated SuperCOS data and the uncalibrated SExtractor analysed image magnitudes for magnitude calibration. The \textsc{Hopcat} calibrated magnitudes are based on SExtractor's  \textsc{mag\_auto} as described in Section \ref{SubSec:OpticalImageData}.

The calculated columns are; optical positions in h:m:s and d:m:s format, \textsc{Hicat-Hopcat} position separation $\Delta$ and axis ratio with the position separation and axis ratio calculated using the following equations:

\begin{equation}
\Delta=2\sin^{-1}\sqrt{\sin^{2}\left(\frac{\Delta dec}{2}\right)+\cos\left(dec_{rad}\right)\cos\left(dec_{opt}\right)\sin^{2}\left(\frac{\Delta ra}{2}\right)}\label{eq:PositionSeparation}\end{equation}

\begin{equation}
axis\: ratio=\frac{semi\: major\: axis}{semi\: minor\: axis}\label{eq:AxisRatio}\end{equation}

where $\Delta$dec=dec$_{radio}$ - dec$_{optical}$ and $\Delta$ra=ra$_{radio}$ - ra$_{optical}$. 

Added columns are SuperCOSMOS image plate numbers, galactic coordinates with $E(B-V)$ extinction values (Schlegel et al.\ 1998), and previously published optical galaxy names and morphologies (NED) for the matched galaxies.

The catalogue contains one line per \textsc{Hopcat} entry. The full version of \textsc{Hopcat} will be made available in {\it Synergy}, the on-line version of the Monthly Notices of the Royal Astronomical Society.  A searchable version of the catalogue is available on-line at:

\begin{center}http://HIPASS.aus-vo.org\end{center}

This database is searchable in a number of ways, including by position and velocity. Returned parameters can be individually selected, along with any of the image products, including detection spectra, on-sky moment maps and position-velocity moment maps. The format of the returned catalogue data can also be chosen, with both HTML and plain text available.

\section{Results}

Starting with the HI radio detections from \textsc{Hicat} and using SuperCOS images, SExtractor image analysis, and published and 6dFGS velocities for match confirmation, we find optical counterparts for 84\% of the HI radio source objects. We use multiple match categories  to identify not only various kinds of matches but the reliability level of the resulting magnitudes. In this section we detail the optically matched galaxy results, analyse the resulting parameters, discuss
the optical properties and possible candidates for dark galaxies.

\subsection{Matched galaxy results }

\begin{figure*}
\centerline{\psfig{file=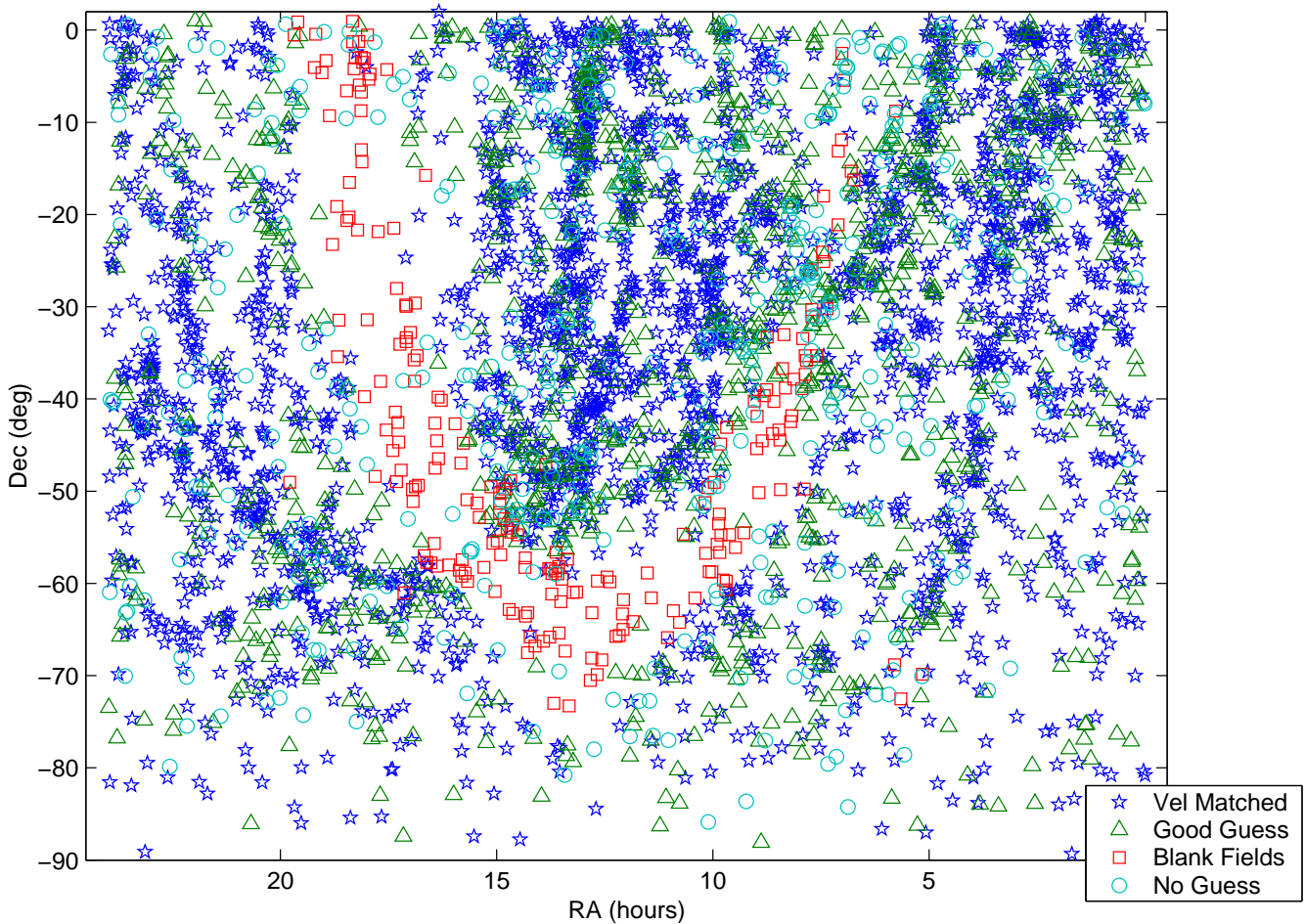,angle=0,width=7in}}
  \caption{\small{Sky plot of all optically matched galaxies.
The insert panel shows the various match categories for all the optical matches.
For a full description of the match categories see Table \ref{Tab:MatchCategoryDescriptions}.
}}
  \label{Fig:All-Southern-SkyHOPCAT}
\end{figure*}

Optical counterparts for 3618 are identified from the 4315 \textsc{Hicat} HI radio sources, including good guesses. Of these 3618 optical counterparts, 972 images contain multiple possible matches, but the best galaxy match is chosen. It is possible that for the multiple velocity matches, the original \textsc{Hicat} velocity could be the average velocity for the whole group. A breakdown of the various categories is listed in Table \ref{Tab:MatchingResults} and a plot of the various categories for the whole southern sky are displayed in  Fig. \ref{Fig:All-Southern-SkyHOPCAT}. It can be seen that, like the sky-plot in Paper I, clear structure
is apparent. Also it should be noted that most of the blank fields are in the galactic plane.

%
%  Table 3.
%
\begin{table*}
\begin{center}
\medskip
\caption{\small{Optical matching results.  For category definitions,
see Table \ref{Tab:MatchCategoryDescriptions}.
}}
\begin{tabular}{lccc}
\hline
Match Category & Category & Number of & Category \\
               & Number   & objects from  &  Percentage of \\
               &          & {\sc Hicat} & {\sc Hicat}\\
\hline 
Velocity Match & 60, 61, 62 & 1798 & 42 \\
Good Guess     & 50, 51, 52 & 848  & 20 \\
Multi Velocity Match & 40, 41, 42 & 714 & 16 \\
Blank Field    & 30         & 216  & 5 \\
No Guess       & 20         & 481  & 11 \\
Multi Good Guess & 10, 11, 12 & 258 & 6 \\
\hline
\end{tabular}
  \label{Tab:MatchingResults}
\end{center}
\end{table*}

%
%  Figure 6.
%

Investigating the results from the matching process can be of great interest and can be used to validate our matching process. The relationship between the HI Radio Integrated Flux and the   \itshape{B$_j$}\upshape \ Apparent Magnitude illustrates whether there is any correlation between the HI radio and optical detections. Fig. \ref{Fig:LogSintVsMag} shows the correlation between these two independent variables. There is a clear trend that shows, the lower the optical magnitude, the larger the HI radio flux. 

When comparing the different match categories, the good guesses tend to reside in the lower left-hand quadrant of the magnitude-flux relationship compared to the velocity matches, with mean values of velocity matched magnitudes=13.8 mag and good guess magnitudes=14.9 mag.  The lower mean 
magnitude for the good guess categories is an expected trend as the smaller  the flux, the fainter the magnitude and hence optical detections and velocity measurements become more difficult.

When we compare our results with the next largest \textsc{Hipass} based study, the \textsc{Hipass} Bright Galaxy Catalogue (BGC, Koribalski et al.\ 2004), several \textsc{Hopcat} optical matches disagree with those from the BGC. Some differences are due to the choice of different galaxies from the same galaxy group. In \textsc{Hopcat} members of a galaxy group are denoted by match categories 10, 11, 12 40, 41, and 42. Other differences in the choice of optical counterparts are due to the different methods used in choosing the matching optical galaxy.  As explained in detail in Section \ref{SubSec:Radio-OpticalMatching} our method involves the visual identification of galaxies using SuperCOS images, SExtractor image analysis of the images to ascertain the galaxy's properties and position, and NED and 6dFGS velocities for cross-checking with \textsc{Hipass} velocities to validate  the optical match. In the case of the BGC however, the optical counterparts were chosen using velocity and position values from NED with some ATCA high resolution HI follow-up observations. 

\subsection{HI Radio-Optical Position and Velocity Separations}
\label{SubSec:HI Radio-Optical Position and Velocity Separations}

The separations between the \textsc{Hicat} HI radio and optical positions for various match categories are shown in Figs. \ref{Fig:PositionSep60s}, \ref{Fig:PositionSep50s}, \ref{Fig:PositionSep40s} and \ref{Fig:PositionSep10s}. The RMS position differences for the velocity matches are $\sigma_{\sc{Hopcat}_{RA}}=0.98$ and $\sigma_{\sc{Hopcat}_{Dec}}=0. 94$ arcmin.
In Paper II (Zwaan et al.\ 2004) fake synthetic point sources are produced to estimate position uncertainty of $\sigma_{\sc{Hipass}_{RA}}$=0.78 and $\sigma_{\sc{Hipass}_{Dec}}$=0.54 arcmin. When comparing the Paper II (Fig. 9) \textsc{Hicat}-synthetic (point) source separation plots with our Figs. \ref{Fig:PositionSep60s}, \ref{Fig:PositionSep50s}, \ref{Fig:PositionSep40s} and \ref{Fig:PositionSep10s}, there is clearly a  higher position separation. The difference between our results  and Paper II's is $\sigma_{dRA}$=0.59 and $\sigma_{dDec}$=0.77 arcmin. We estimate this difference according to:\begin{equation} \sigma_{dRA}^{2}=\sigma_{\sc{Hopcat}_{RA}}^{2}-\sigma_{\sc{Hipass}_{RA}}^{2}
\label{eq:Hicat-HopatSigmaPosDiff}\end{equation} This difference might be accounted for by the difference in the objects being compared, i.e. synthetic point sources compared with our extended
objects. One reason for this uncertainty is the difference between the central position for the HI radio and optical detected objects. Optically detected galaxies tend to be symmetric with a central peak luminosity whereas HI detections are more likely to be asymmetric and the peak flux density
may not be central. This difference in symmetry and position for a single object's peak flux density and luminosity detected positions may cause the larger separation. 

\begin{figure}
\centerline{\psfig{file=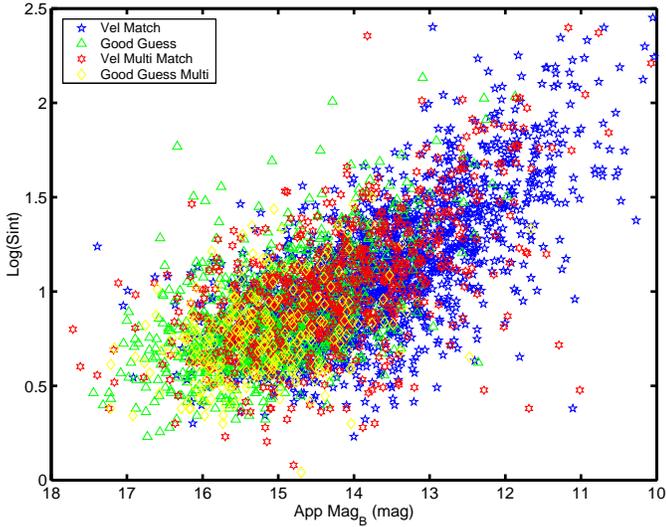,angle=0,width=3.5in}}
  \caption{\small{Log (integrated HI radio flux) versus  \itshape{B$_j$}\upshape \ apparent magnitudes for the optically matched \textsc{Hicat} sources. Match categories are listed in the insert.  This shows there is a clear trend that, the brighter the optical magnitude, the larger the HI radio flux.
}}
  \label{Fig:LogSintVsMag}
\end{figure}

\begin{figure}
  \centerline{\psfig{file=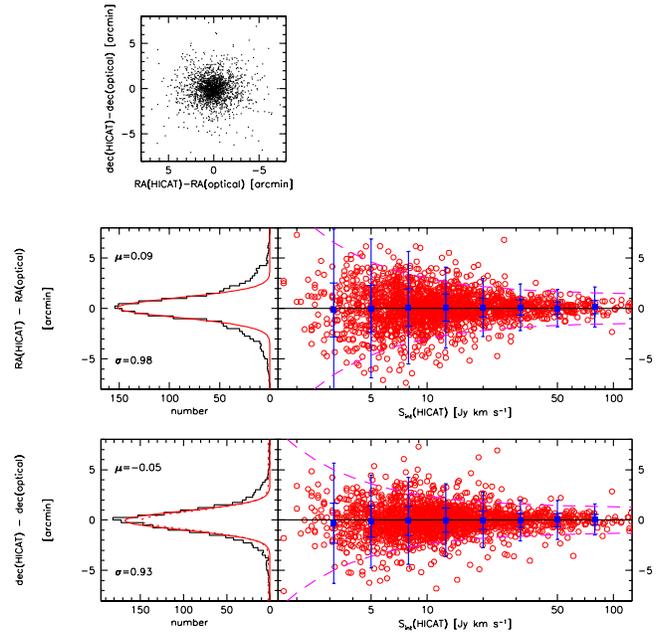,angle=0,width=3.5in}}
  \caption{\small{Separations between the \textsc{Hicat} and \textsc{Hopcat} 
optical positions for velocity matched galaxies (match categories 60, 61 and 62). The top panel shows the difference between the HI radio and optical RA and Dec positions. For the middle and bottom panels, the left-hand panels are histograms of the HI radio and optical position differences fitted by Gaussian profiles with the parameters indicated in the top left hand corner. The right-hand panels show the relationship between the Integrated flux and the position separation for RA and Dec respectively. See Table \ref{Tab:MatchCategoryDescriptions} for full match category descriptions.
}}
  \label{Fig:PositionSep60s}
\end{figure}

\begin{figure}
  \centerline{\psfig{file=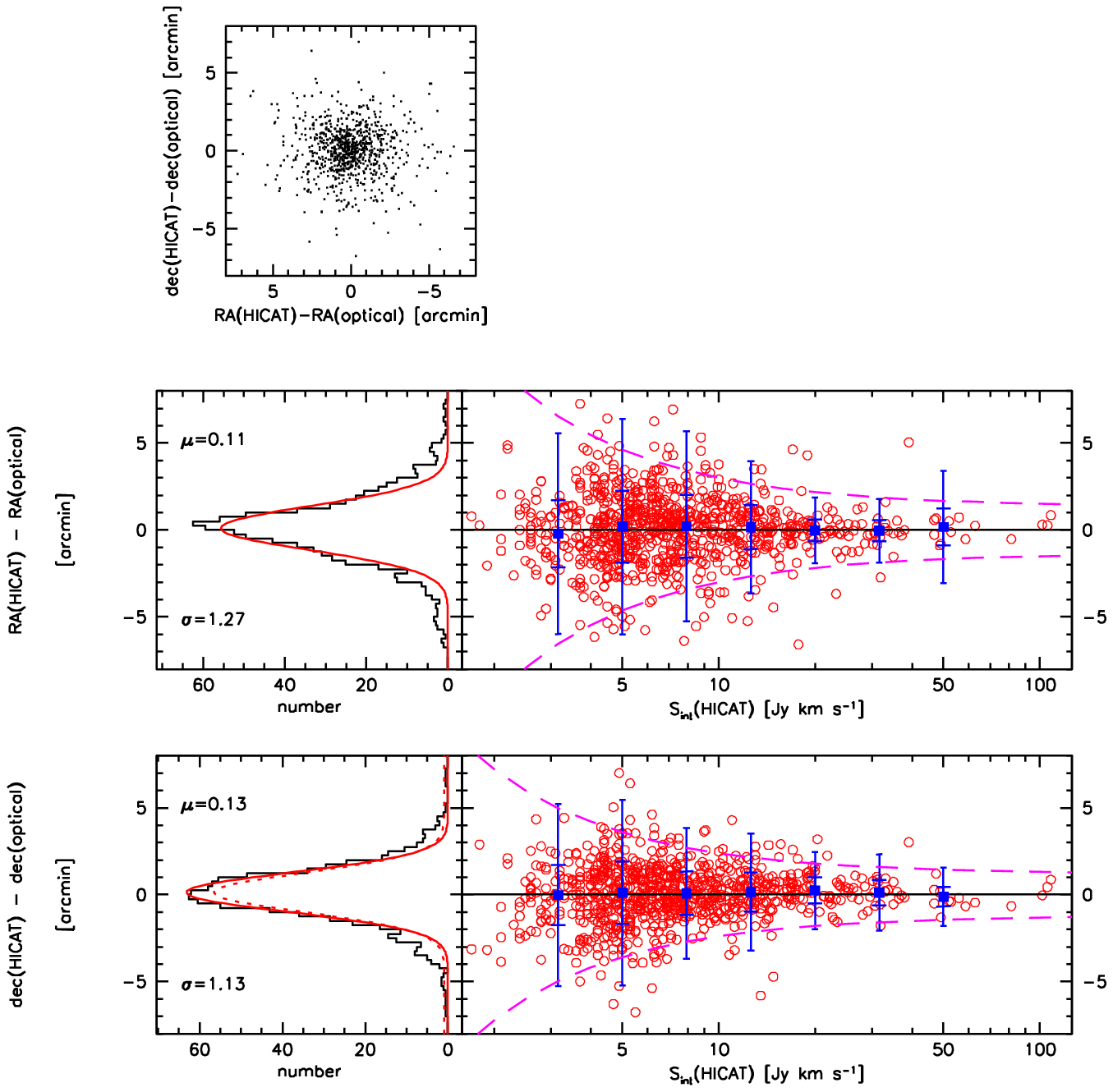,angle=0,width=3.5in}}
  \caption{\small{Separations between \textsc{Hicat} and \textsc{Hopcat} optical positions for good guess matched galaxies (match categories 50, 51 and 52). See Fig. \ref{Fig:PositionSep60s} for panel descriptions and Table \ref{Tab:MatchCategoryDescriptions} for full match category descriptions.
}}
  \label{Fig:PositionSep50s}
\end{figure}

\begin{figure}
  \centerline{\psfig{file=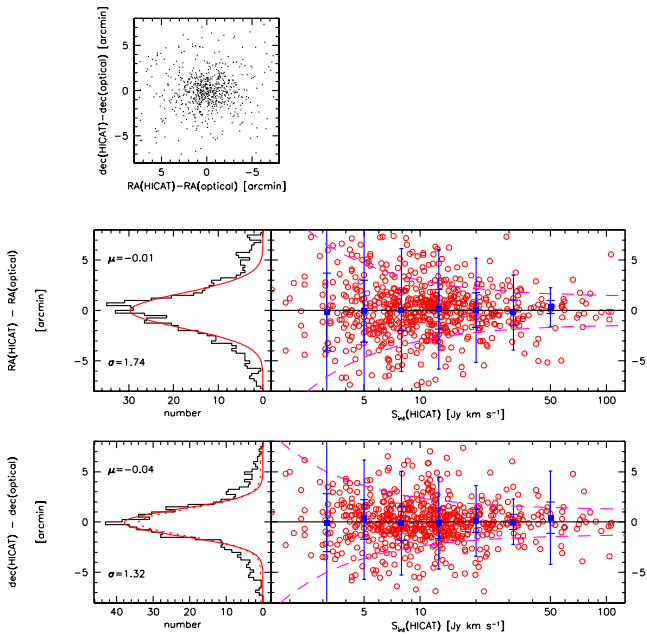,angle=0,width=3.5in}}
  \caption{\small{Separations between \textsc{Hicat} and \textsc{Hopcat} optical positions for multiple velocity matched galaxies (match categories 40, 41 and 42). See Fig. \ref{Fig:PositionSep60s} for panel descriptions and Table \ref{Tab:MatchCategoryDescriptions} for full match category descriptions.
}}
  \label{Fig:PositionSep40s}
\end{figure}

\begin{figure}
  \centerline{\psfig{file=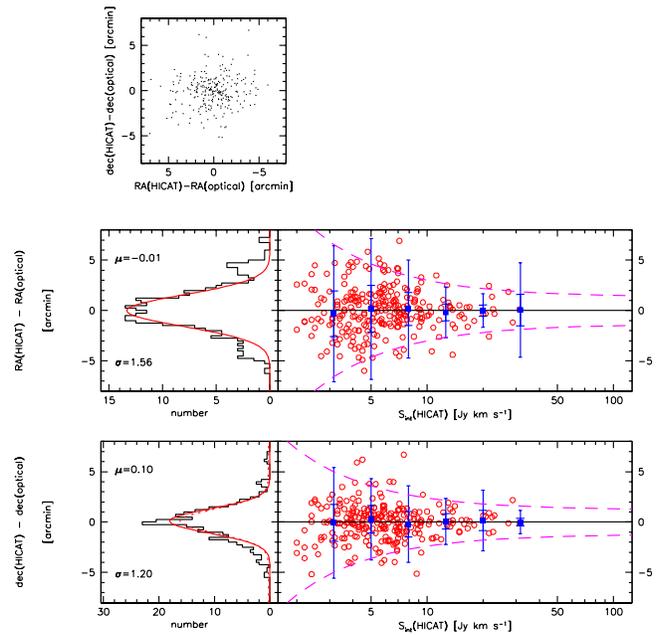,angle=0,width=3.5in}}
  \caption{\small{Separations between \textsc{Hicat} and \textsc{Hopcat} 
optical positions for multiple good guess matched galaxies (match categories 10, 11 and 12). 
See Fig. \ref{Fig:PositionSep60s} for panel descriptions and Table \ref{Tab:MatchCategoryDescriptions} for full match category descriptions.
}}
  \label{Fig:PositionSep10s}
\end{figure}

Fig. \ref{Fig:PositionSep60s} shows the velocity matches that are verified by independent published velocities, Fig. \ref{Fig:PositionSep50s} the good guesses,  Fig. \ref{Fig:PositionSep40s} the multiple velocity matches and Fig. \ref{Fig:PositionSep10s} the multiple good guess matches. Note that there are fewer good guesses in the upper limits of the integrated flux. This is understandable since the high flux galaxies are more likely to have previously observed bright optical counterparts (Fig. \ref{Fig:LogSintVsMag}) with measured velocities. This means the larger $\sigma$ for the good guesses is expected as this category matches mainly the fainter galaxies. We can deduce from the similar spread in the velocity matched and good guess plots that the good guesses are indeed acceptable optical counterparts for the \textsc{Hicat} HI radio sources. 

The separation between HI radio and optical positions for the other two match categories, multiple velocity and multiple good guess matches, are shown in Figs. \ref{Fig:PositionSep40s} and \ref{Fig:PositionSep10s} respectively. For Fig. \ref{Fig:PositionSep40s}, there is a larger spread for multiple velocity matches position separations than for the velocity matches. These matches are from images that contain multiple galaxies with similar velocities. The galaxy with the closest independent published velocity value is chosen, although the original HI detection position could be either the highest or the total velocity of the whole group.  This effect may distort the position separation where we use velocity for matching.  We note that the good guess multiple galaxy matches also have a larger $\sigma$ than the good guess matches (Fig. \ref{Fig:PositionSep10s}).  We have high resolution 21cm radio observations for 40 of these velocity and good guess multiple galaxy matches and more observations are planned. From these extra observations we will be able to determine statistically the quality of these match categories.

For the optically matched galaxies, velocity matches (6's), multiple velocity matches (4's), good guesses (5's) and multiple good guesses (1's), any HI radio-optical galaxy separation greater than 7.5 arcmin are not considered a reliable match. Matches beyond this limit, corresponding to 0.6 per cent of the velocity matches, we use with caution.

%
%  Figure 7.
%

%
%  Figure 8.
%

%
%  Figure 9.
%

% Figure 10.
%
%
%  Figure 11.
%

\begin{figure}
\centerline{\psfig{file=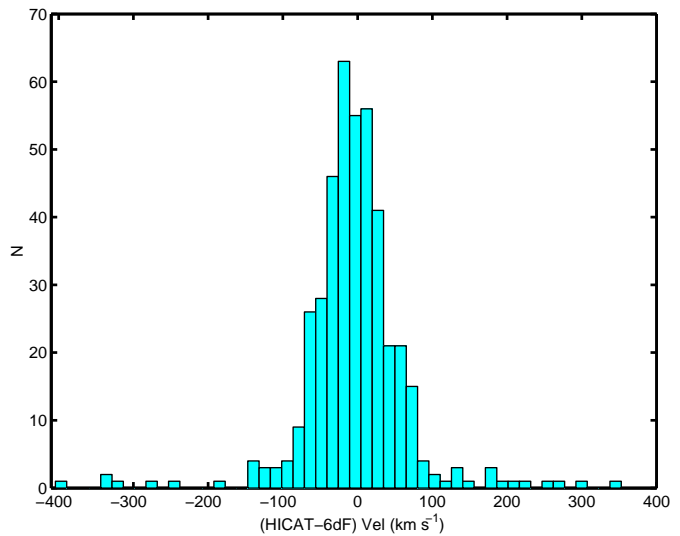,angle=0,width=3.5in}}
  \caption{\small{Difference between the \textsc{Hicat} HI radio and
the matched galaxy velocities from 6dFGS. A velocity match is where the 
\textsc{Hicat} and NED and/or 6dFGS velocity difference is within $\Delta velocity=400\:km\:s^{-1}$.
}}
  \label{Fig:VelDiff}
\end{figure}

When confirming the \textsc{Hicat} velocities using published and 6dFGS velocities, some velocity difference cut-off is needed where galaxies with velocities greater than the cut-off are not considered a velocity confirmed optical match. Fig. \ref{Fig:VelDiff} shows the distribution of galaxies that have velocity matched confirmation. From this we have determined the cut off value for confirmed velocity matches as 400 km s$^{-1}$corresponding to 5$\sigma$. The majority of velocity confirmed optical matches have a velocity difference of $<$ 100 km s$^{-1}$. 

%
%  Figure 12.
%

\subsection{Candidate Dark Galaxies }

\begin{figure}
\centerline{\psfig{file=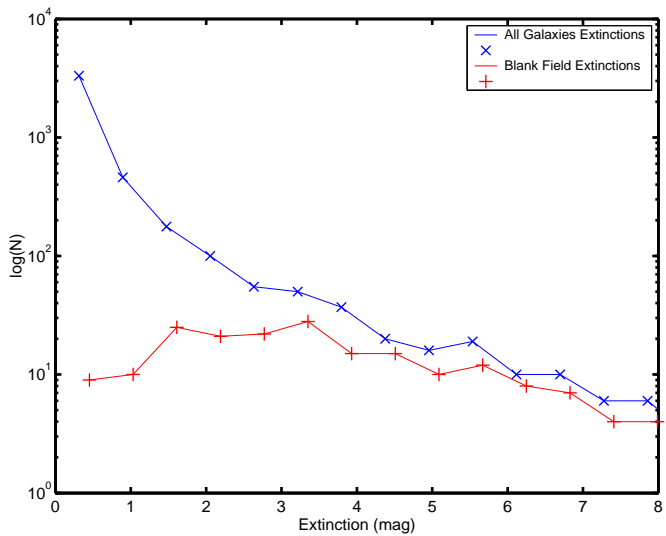,angle=0,width=3.5in}}
  \caption{\small{The number of blank fields compared to all \textsc{\small Hopcat} galaxies as
a function of their $A_{B_j}$ extinction values. An $A_{B_j}$ extinction cut-off of 1 mag is taken to be the upper limit of $A_{B_j}$ extinction when looking at blank fields to find possible candidates for isolated dark galaxies.
}}
  \label{Fig:FracBFforExt}
\end{figure}

As discussed in Section \ref{Sec:Introduction} the \textsc{Hicat} data may reveal whether isolated dark galaxy exist or not. From our matching process 216 are classed as blank fields. These fields contain no visible optical galaxies. The selection criteria used to search for dark galaxy candidates are extinction cut-off and the blank field category. \textsc{Hopcat} contains $E(B-V)$ extinction values but as we are using $B_j$ images, a correction must be applied, $A_{B_j}=4.035\times E(B-V)$ (Schlegel et al., 1998). We compare the $A_{B_j}$ extinction distribution for all the \textsc{Hopcat} galaxies and the blank field category (Fig. \ref{Fig:FracBFforExt}), and take an $A_{B_j}$ extinction cut at 1 mag as, beyond this extinction, optically faint galaxies will be dust obscured. 

A total of 3692 galaxies have an $A_{B_j}$ extinction $<$ 1 mag, with only 13 galaxies also in the blank field category. From these, 11 are found to be in over-crowded fields. One object, HIPASSJ1351-47, on close inspection does have a faint optical counterpart.  This object is listed in Banks et al. (1999) as a new galaxy in the Centaurus A group. The final dark galaxy candidate is not a real HI detection. As described in Zwaan et al. (2004), many \textsc{Hicat} detections have been re-observed with the Parkes Narrow band system to assess the reliability of \textsc{Hicat}.  Marginal sources confirmed as false detections were removed from the catalogue, however not all sources were re-observed.  This final dark galaxy candidate, HIPASSJ1946-48, has recently been confirmed as a false detection.

Our conclusion is that from the 4315 HI radio detections in \textsc{Hicat}, of which 3692 galaxies have an $A_{B_j}$ extinction $<$ 1 mag, no isolated dark galaxies have been found.

%
%  Figure 13.
%

\subsection{Mass-to-light ratio.}

\begin{figure*}
\centerline{\psfig{file=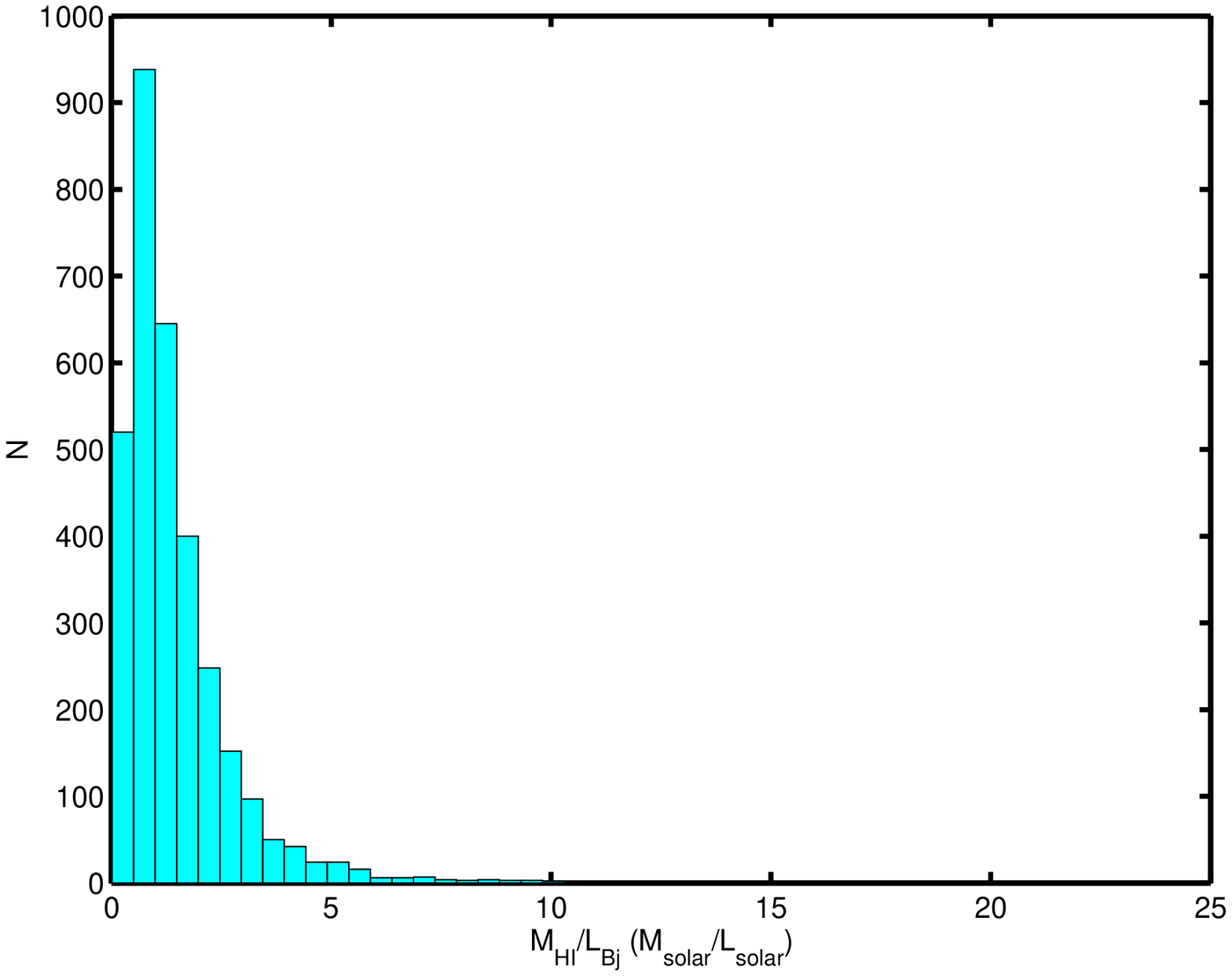,angle=0,width=3.5in}
\psfig{file=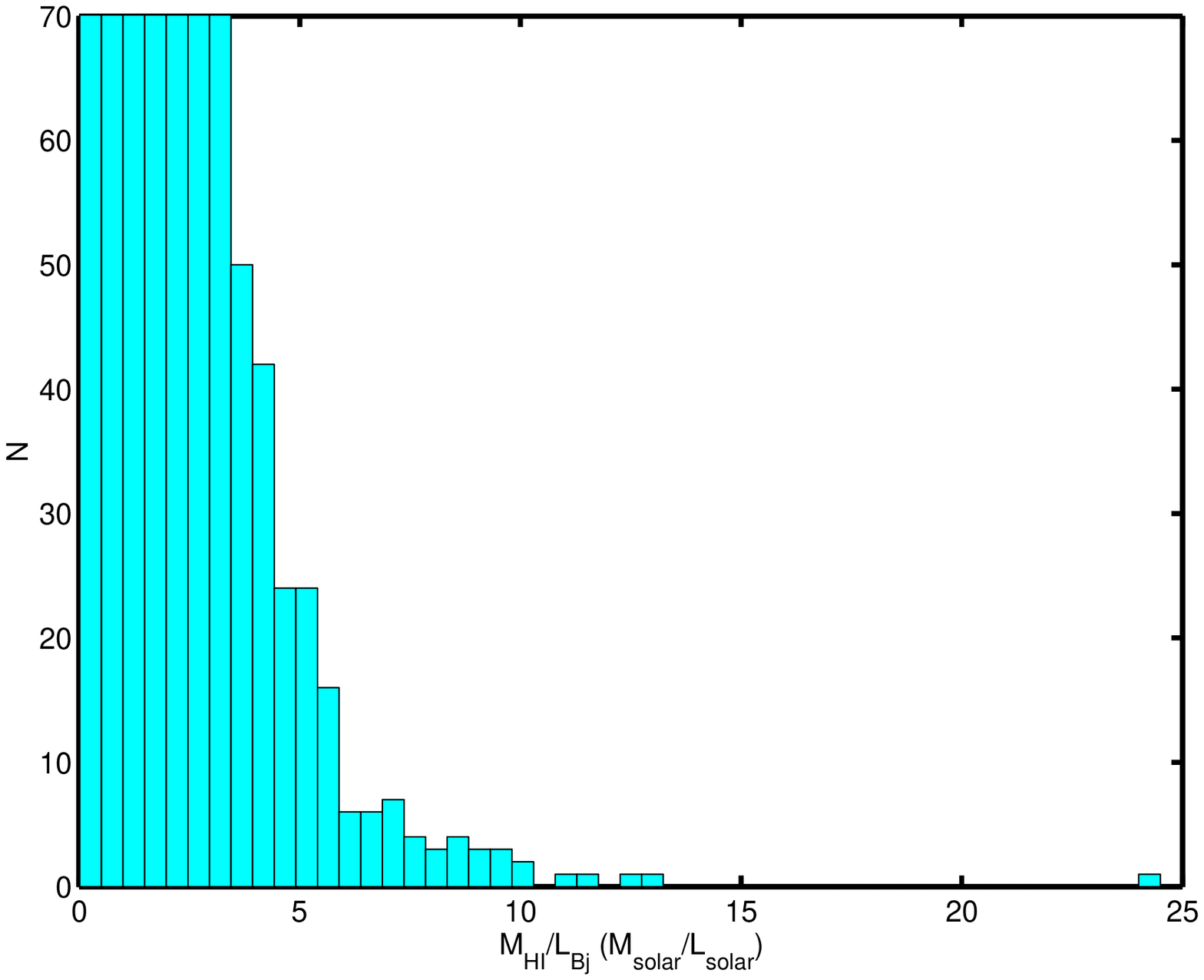,angle=0,width=3.5in}}
  \caption{\small{HI Mass to $B_j$ Light ratio
plots. Left: For all matched galaxies with $A_{B_j}$ extinction $<1$mag. 
Right: Enlargement showing the rapid decrease in the number of galaxies as the mass-to-light ratio increases out to 13 $M_{HI\odot}/L_{Bj\odot}$ and one high mass-to-light ratio LSB galaxy where the image analysis process is not able to calculate an SEx ellipse to represent the total area of the galaxy, hence underestimating the luminosity.
}}
  \label{Fig:MassToLLightRatioPlots}
\end{figure*}

We investigate the mass-to-light ratio using the HI mass and $B_j$ band luminosity ratio for optically matched \textsc{Hicat} HI radio sources. The original $B_j$ apparent magnitudes are extinction corrected and have an $A_{B_j}$ extinction $<$ 1 mag (Fig. \ref{Fig:MassToLLightRatioPlots}). The majority of these galaxies have mass-to-light ratios less than 5 $M_{\odot}/L_{\odot}$. The number of galaxies rapidly decreases with increasing mass-to-light ratio up to 13 $M_{\odot}/L_{\odot}$. The one galaxy with an higher mass-to-light ratio, HIPASSJ1227+01 at 24.5 $M_{\odot}/L_{\odot}$, is a LSB galaxy where the image analysis process is not able to calculate a SEx ellipse that represents the total area of the galaxy, hence the luminosity is underestimated. This object, was first discovered by Giovanelli \& Haynes (1989) and has a measured mass-to-light ratio of 10.3 $M_{HI\odot}/L_{B\odot}$ in  Salzer et al. (1991).

When comparing our mass-to-light ratios with a subset of \textsc{Hipass} of the South celestial cap region (Kilborn et al.\ 2002), though a much smaller sample, the same mass-to-light relationship
is found. Again most galaxies have a ratio less than 5 $M_{\odot}/L_{\odot}$ with quickly decreasing galaxy numbers for increasing mass-to-light ratios up to 13 $M_{\odot}/L_{\odot}$ and few galaxies with greater than 15 $M_{\odot}/L_{\odot}$.

%
%  Figure 14.
%

\section{Summary}

Our catalogue, \textsc{Hopcat}, matches the \textsc{Hipass} Catalogue, \textsc{Hicat} entries with their optical counterparts and is the largest catalogue ever produced that optically identifies HI radio detected sources. We identify optical counterparts for 84\% of the HI radio sources. Of these 20\% have multiple possible matches. No guess is possible for 11\% and 5\% are blank fields. Most of the blank fields are in crowded fields along the galactic plane or have high $E(B-V)$ extinction. For a full description of \textsc{Hopcat's } columns see Table \ref{Tab:HopcatColumns}

Using an extinction cut of  $A_{B_j}$ $<$ 1 mag and the blank field category, only 13 HI radio sources remain in our dark galaxy search. Of these, 12 are eliminated due to over-crowded fields or by Parkes Narrow-band follow-up observations. The remaining one, on close inspection, does have a faint optical counterpart. Our conclusion: no isolated dark galaxies exist within the limits of the \textsc{Hipass} survey.

All \textsc{Hicat} data including optical counterparts is available on-line at:

\begin{center}http://HIPASS.aus-vo.org\end{center}

We encourage other researchers to make use of this database. For optimum utility, researchers also need to be aware of the completeness, reliability and accuracy of the measured parameters. These are described in detail in Paper II (Zwaan et al.\ 2004) and the optical data in this paper. In particular, users are reminded of the uncertainties associated with the various 2 digit match categories. These define the choice of the optical match; velocity match (6's), good guess (5's) etc and the quality of the match; good quality match (0), poor photometry (1), and poor segmentation (2). Users are also encouraged to be familiar with the full processing of \textsc{Hipass} data (Barnes et al.\ 2001; Meyer et al.\ 2004).

\section*{Acknowledgments}
The Multibeam system is funded by the Australia Telescope National Facility (ATNF) and an Australian Research Council grant. The collaborating institutions are the Universities of Melbourne, Western Sydney, Sydney, Cardiff, the Research School of Astronomy and Astrophysics at Australian National University (RSAA), Jodrell Bank Observatory and the ATNF. The Multibeam receiver and correlator is designed and built by the ATNF with assistance from the Australian Commonwealth Scientific and Industrial Research Organization Division of Telecommunications and
Industrial Physics. The low noise amplifiers used for \textsc{Hipass} were provided by Jodrell Bank Observatory through a grant from the UK Particle Physics and Astronomy Research Council. The Multibeam Survey Working Group is acknowledged for its role in planning and executing the \textsc{Hipass} project. This work makes use of the \textsc{aips}++, \textsc{miriad} and \textsc{Karma} software packages. 

We would also like to acknowledge the assistance of the 6dF Galaxy Survey  conducted primarily by The Anglo-Australian Observatory, and with support from RSAA and the Wide-field Astronomy Unit of the University of Edinburgh. Also to Danielle Parmenter who helped with the first full HI radio/optical galaxy matching.

This research has made use of the NASA/IPAC Extragalactic Database which is operated by the Jet Propulsion Laboratory, California Institute of Technology, under contract with the National Aeronautics and Space Administration. 

MTD is supported through a University of Queensland Graduate School Scholarship.  KAP acknowledges support from an EPSA University of Queensland Research Fellowship and UQRSF grant. This work has also been supported by a University of  Queensland Research Development Grant and by DP and LIEF grants from the Australian Research Council.

%
%
%  Table 6 Appendix.
%
\begin{table*}
\begin{center}
\medskip
\caption{\small{\textsc{Hipass} parameter description}}
\begin{tabular}{lllll}
\hline
Column No.&Parameter & Database Name        & Units         & Description \\
\hline 
1&ID &    ID &    N/A    &  \\
2&Hipass Name &   hipass\_name &   N/A &   Names are of the form Hipass JXXXXYY[a-z],\\ 
 & & & &                                   where is the unrounded source RA in hrs and  \\
 & & & &                                 min, and YY is the unrounded source declination \\
 & & & &                                 in degrees. An additional letter a-z is added where \\
 & & & &                                 necessary to distinguish sources.\\
3&HI radio Right Ascension & RA &    h:m:s & Right ascension (J2000 hexadecimal format) \\
4&HI radio Declination &     Dec &   d:m:s & Declination (J2000 hexadecimal format) \\
5&HI Velocity $^{\ddagger}$  & vel\_mom & km s$^{-1}$& Flux weighted velocity average between manually\\
 & & & &																						specified minimum ($v_{lo}$) \& maximum ($v_{hi}$)\\
 & & & &																						 profile velocity\\
6&HI Velocity Width & width\_50max & km s$^{-1}$ & Difference of velocity at which profile reaches \\
 & & & &                                   50 per cent of peak flux density\\
7&HI Peak Flux Density &     Sp &     Jy &    Peak flux density of profile \\
8&HI Integrated Flux &  Sint &  Jy km s$^{-1}$ &  Integrated flux of source (within region $v_{lo}$ \& $v_{hi}$ \& \\
 & & & &                                           box size) \\
9&Semi-major Axis &  A\_IMAGE &  pixel &  Profile RMS along major axis \\
10&Semi-minor Axis &  B\_IMAGE &  pixel &  Profile RMS along minor axis \\
11&Position Angle &  THETA\_IMAGE &   deg & Position angle\\
12& 6dF Velocity $^{\ddagger}$ &    Velocity\_6dF &  km s$^{-1}$ &   Velocity from 6dF Data \\
13&NED Velocity $^{\ddagger}$  &   Velocity\_Ned &  km s$^{-1}$ &   Velocity From NED Data \\
14&Optical Match  Category $^{\dagger}$ &    Class(Matched\_choice) & N/A &   Our category matching choice \\
 & & &  &  \\
15&Optical Right Ascension  &  SExRA & deg & Right ascension (J2000)  \\
& (Degrees) & & &\\
16&Optical Declination (Degrees) & SExDec &  deg & Declination (J2000) \\
17&Optical Right Ascension &  Matched\_RA\_H:M:S &   h:m:s & Right ascension (J2000)\\
& (h:m:s) & & & \\
18&Optical Declination (d:m:s) &   Matched\_Dec\_D:M:S &     d:m:s & Declination (J2000) \\
19&Calibrated  \itshape{B$_j$}\upshape \ Magnitude & $B_j$\_Mag\_AUTO &  mags &  $B_j$ band calibrated magnitude based on SExtractor\\
&  &Calibrated & &                                     Kron-like elliptical aperture magnitude \\
20&Calibrated \itshape{R}\upshape \ Magnitude &  Red\_Mag\_AUTO &  mags &  $R$ band calibrated magnitude based on SExtractor\\
&  & \_Calibrated & &                                     Kron-like elliptical aperture magnitude \\
21&Calibrated \itshape{I}\upshape \ Magnitude & \itshape{I}\upshape \_Mag\_AUTO & mags & \itshape{I}\upshape \ band calibrated magnitude based on SExtractor \\
&  &\_Calibrated & &                                Kron-like elliptical aperture magnitude \\
22&SuperCOSMOS  \itshape{B$_j$}\upshape \ Plate Number  & BluePlateNumber &  N/A &   $B_j$ band SuperCOSMOS original image plate number\\
&  & & & \\
23&SuperCOSMOS \itshape{R}\upshape \ Plate Number  &  RedPlateNumber &   N/A &   $R$ band SuperCOSMOS original image plate number\\
&  & & & \\ 
24&SuperCOSMOS \itshape{I}\upshape \ Plate Number &    IPlateNumber &  N/A & \itshape{I}\upshape \ band SuperCOSMOS original image plate number\\
25&l &     HicatExtl &     degrees &  Galactic longitude \\
26&b &     HicatExtb &     degrees &  Galactic latitude \\
27&Extinction $E(B-V)$ &    HicatExt &      mag &   Extinction values ({Schlegel} et al., 1998) \\
28&RA HI radio-Optical Position &  DiffRaArcmin &  arcmin &  Right Ascension Position separation Hicat-Hopcat \\
& Separation & & & (radio-optical) \\
29&Dec HI radio-Optical Position  & DiffDecArcmin & arcmin &   Declination Position separation Hicat-Hopcat \\
& Separation & & & (radio-optical) \\
30&Galaxy HI radio-Optical Position &  GalSepArcmin &  arcmin & Galaxy Position separation Hicat-Hopcat \\
& Separation & & &  (radio-optical)\\
31&Axis Ratio &    AxisRatio &    N/A &   Semi-major Axis/Semi-minor Axis \\
32&Published Optical GalaxyName & Galaxy Name &   N/A &   Published Optical Names from NED\\           
33&Published Morphology & Morphology &   N/A &   Published Morphology from NED\\                  
\hline
\end{tabular}
\label{Tab:HopcatColumns}
$^{\dagger}$ 2 digit optical match category\\
First Digit: 6=velocity match; 5=Good Guesses; 4=multiple velocity match; 3=blank field; 2=no guess; 
1=multiple good guess\\
Second digit: 0 = good photometry and segmentation, 1 = Poor photometry, 2 = poor segmentation\\
$^{\ddagger}$  All velocities are cz \&  Heliocentric

\end{center}
\end{table*}

\end{document}